\documentclass[10pt]{article}
\usepackage{amsmath}
\usepackage{amssymb}
\usepackage[dvips]{graphicx}
\usepackage[dvips]{epsfig}
\usepackage{color}

 \advance \textwidth by 0.1cm
 \advance \textheight by 0.06cm
\def\##1{{\bf #1}}
\def\=#1{\underline{\underline #1}}
\def\4#1{\underline{\underline{\underline{\underline #1}}}}

\def\.{\mbox{ \tiny{$^\bullet$} }}

\def\r#1{(\ref{#1})}

\def\epso{\epsilon_{\scriptscriptstyle 0}}
\def\lambdao{\lambda_{\scriptscriptstyle 0}}
\def\muo{\mu_{\scriptscriptstyle 0}}
\def\ko{k_{\scriptscriptstyle 0}}
\def\etao{\eta_{\scriptscriptstyle 0}}

\def\eps{\epsilon}

\def\epsmet{\eps_m}

\def\Einc{{\mathbf E}_{inc}({\bf r})}
\def\Erefl{{\mathbf E}_{ref}({\bf r})}
\def\Etr{{\mathbf E}_{tr}({\bf r})}

\def\Hinc{{\mathbf H}_{inc}({\bf r})}
\def\Hrefl{{\mathbf H}_{ref}({\bf r})}
\def\Htr{{\mathbf H}_{tr}({\bf r})}

\def\rp{r_p}

\def\tp{t_p}

\def\ux{{\mathbf{u}}_x}
\def\uy{{\mathbf{u}}_y}
\def\uz{{\mathbf{u}}_z}

\def\utau{{\mathbf{u}}_\tau}
\def\un{{\mathbf{u}}_n}
\def\ub{{\mathbf{u}}_b}

\def\vkap{\varkappa}

\begin{document}

\begin{center}
{\large {\bf Morphological influence on surface--wave
propagation at the planar interface of a metal film and a
 columnar thin film}}

Akhlesh Lakhtakia$^1$ and John A. Polo, Jr.$^2$

{\small $^1$\emph{Department of Engineering Science \& Mechanics,
Pennsylvania State University, \\ University Park, PA 16802, USA}}\\
{\small $^2$\emph{Department of Physics and Technology,
Edinboro University of Pennyslvania,\\
Edinboro, PA  16444, USA}}
\end{center}
\noindent{\small
The selection of a higher
vapor deposition angle when growing a columnar thin film (CTF) leads to surface--wave propagation at a
planar metal--CTF interface
with phase velocity of lower magnitude and shorter propagation
range. Acordingly, a higher angle of plane--wave incidence is required to
excite that surface wave in a modified Kretschmann configuration. \copyright\,
Anita Publications. All rights reserved.
}\\

\noindent{\bf 1 Introduction}

An enormous body of  literature on
the propagation of electromagnetic waves  localized to the planar interfaces of bulk metals and bulk dielectric materials has gathered during the last hundred years
 [1-3]. Such a wave is associated in quantum parlance with
\emph{surface plasmon polaritons}, resulting from the interaction of photons in the dielectric material
and electrons in the metal. The quantum term is parsed as follows: the entities are localized to a
\emph{surface}; a \emph{plasmon} is a collective excitation of electrons; and the dielectric 
material is polarized because of interaction with photons,
thereby giving rise to the noun
\emph{polariton}. In the language of classical electromagnetics, the surface waves are $p$--polarized, not $s$--polarized.

Although the dielectric material is usually considered to be  isotropic and homogeneous,
anisotropic dielectric materials (such as crystals) have been incorporated in
surface--wave research as well [4,5]. As is known well,
assemblies of parallel nanowires called columnar thin films (CTFs) can be
used in optics in lieu of crystalline dielectric materials [6]. Provided the wave vector
of an electromagnetic planar wave is oriented parallel to the morphologically significant
plane of a CTF, a distinction between $p$-- and $s$--polarization states
can be easily made [7]; hence, it can be conjectured immediately that
surface waves can propagate on the
planar interface of a homogeneous metal and a CTF.

\begin{center}
\begin{figure}[!htb]
\centering \psfull
\epsfig{file=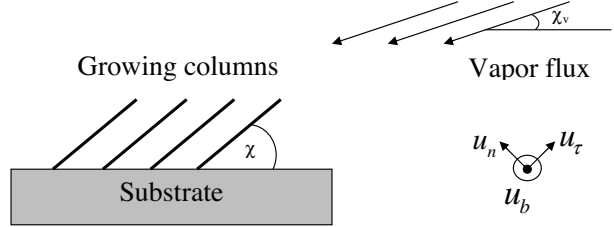, width=8cm}
%\bigskip
\caption{
{\small Schematic of the growth of a columnar thin film. The vapor flux
is directed at an angle $\chi_v$, whereas columns grow at an angle
$\chi\geq\chi_v$. The unit vectors $\utau$, $\un$, and $\ub$ are also
shown.
\label{Fig:geometry}
}}
\end{figure}
\end{center}

Surface--wave propagation (SWP) on a planar metal--CTF interface is bound to be
affected by the morphology of the CTF. Columnar thin films    fall under the general banner of
biaxial dielectric materials for optical purposes.
These thin films are grown by physical vapor
deposition, whereby vapor from a source boat in an evacuated chamber is directed at angle
$\chi_v\in\left(0,\pi/2\right]$ towards a planar substrate, as shown in Figure~\ref{Fig:geometry}. Under the right
conditions, parallel columns of the evaporant species grow on the substrate tilted at an angle $\chi\geq\chi_v$.
The CTFs are composed of multimolecular clusters  with $\sim3$~nm diameter  which, in turn, are clustered
in a fractal--like nature eventually forming columns with $\sim 100$-nm  cross--sectional diameter,  depending on the evaporant species and the deposition
conditions.
Once the evaporant species and the deposition conditions have
been chosen, the vapor incidence angle $\chi_v$ can be used to alter the morphology of a CTF significantly enough as to have optical consequences.
Indeed, at visible frequencies and lower, a CTF may be regarded as a linear, orthorhombic
continuum whose relative permittivity dyadic can be controlled
by proper selection of  $\chi_v$ [8].

Our aim in this communication is to establish the influence of the CTF
morphology, as captured by the vapor incidence angle, on SWP at a planar
metal--CTF interface.
Section~2
provides the derivation of the SWP wavenumber at a planar metal--CTF
interface, and Section~3 contains the solution of a boundary--value
problem to excite a surface wave in a Kretschmann configuration [9,10] modified
for practical considerations.
An $\exp(-i\omega t)$ time--dependence is implicit, with $\omega$
denoting the angular frequency. The free--space wavenumber, the
free--space wavelength, and the intrinsic impedance of free space are denoted by $\ko=\omega\sqrt{\epso\muo}$,
$\lambdao=2\pi/\ko$, and
$\etao=\sqrt{\muo/\epso}$, respectively, with $\muo$ and $\epso$ being  the permeability and permittivity of
free space. Vectors are in boldface, dyadics underlined twice;
column vectors are in boldface and enclosed within square brackets, while
matrixes are underlined twice and similarly bracketed. Cartesian unit vectors are
identified as $\ux$, $\uy$ and $\uz$.\\

%%%%%%%%%%  Figure 1 begins %%%%%%%%%%%%
\smallskip
%\smallskip
%%%%%%%%%%  Figure 1 ends  %%%%%%%%%%%%

\noindent{\bf 2 SWP at Planar Metal--CTF Interface}

Let the region $z>0$ be occupied by a CTF and the region $z<0$ by a
metal of relative permittivity $\epsilon_m$. The relative permittivity dyadic of the CTF may be stated
as [7]
\begin{equation}
\=\eps_{\,ctf}=\epsilon_a\,\un\un+\epsilon_b\,\utau\utau+\epsilon_c\,\ub\ub\,,\label{Eqn:epsCTF}
\end{equation}
where $\epsilon_{a,b,c}$ are the principal relative
permittivity scalars and the unit vectors
\begin{equation}
\un= -\ux\sin\chi+\uz\cos\chi\,,\quad
\utau=\ux\cos\chi+\uz\sin\chi\,,\quad
\ub=-\uy\,.
\end{equation}
All four constitutive quantities~---~$\epsilon_{a,b,c}$ and the column inclination angle $\chi\in\left(0,\pi/2\right]$~---~depend on the evaporant species
and the vapor incidence angle $\chi_v$.  The $xz$ plane, containing the
unit vectors $\un$ and $\utau$, is the morphologically significant plane of the
CTF.

In order to preserve the independence of the $p$-- and the $s$--polarization states,
as stated in Section~1,
the wave vectors in both half--spaces must not have a component
orthogonal to the morphologically significant plane. Accordingly,
the $p$--polarized electromagnetic field phasors in the metal half--space
are given by
\begin{equation}
\displaystyle{
\left.
\begin{array}{l}
\#E(\#r)=A_m\left(\ux -\frac{i\vkap}{q_m}\uz\right) \exp\left[i\ko\left(\vkap x-iq_mz\right)\right]\\[5pt]
\#H(\#r)=A_m\,\etao^{-1}\,\frac{i\epsilon_m}{q_m}\,\uy\,
 \exp\left[i\ko\left(\vkap x-iq_mz\right)\right]
 \end{array}\right\}\,,\quad z \leq 0
 }\,,
\end{equation}
where $A_m$ is a complex--valued scalar of finite magnitude, $q_m=+\sqrt{\vkap^2-\epsilon_m}$, and $\vkap\ko\ux$ may be regarded as the wave vector of the
surface wave.  We must
have ${\rm Re}\left[q_m\right]>0$ for SWP.
The $p$--polarized electromagnetic field phasors in the CTF
half--space are given by [7]
\begin{equation}
\displaystyle{
\left.
\begin{array}{l}
\#E(\#r)=A_c\left[\ux+
\frac{i\vkap q_c-(\epsilon_a-\epsilon_b)\sin\chi\cos\chi}{\vkap^2-(\epsilon_a\cos^2\chi+\epsilon_b\sin^2\chi)}\,\uz\right]
 \exp\left[i\ko\left(\vkap x+iq_cz\right)\right]\\[5pt]
\#H(\#r)=A_c\,\etao^{-1}\left[
\frac{-iq_c(\epsilon_a\cos^2\chi+\epsilon_b\sin^2\chi)+\vkap(\epsilon_a-\epsilon_b)\sin\chi\cos\chi
}{\vkap^2-(\epsilon_a\cos^2\chi+\epsilon_b\sin^2\chi)}\right]\uy\,
 \exp\left[i\ko\left(\vkap x+iq_cz\right)\right]
 \end{array}\right\}\,,\quad z \geq 0
 }\,,
\end{equation}
where $A_c$ is a complex--valued scalar of finite magnitude,   and $q_c$
must be obtained by solving the quadratic equation
\begin{equation}
\vkap^2(\epsilon_a\sin^2\chi+\epsilon_b\cos^2\chi)
-2i\vkap q_c(\epsilon_a-\epsilon_b)\sin\chi\cos\chi
-q_c^2(\epsilon_a\cos^2\chi+\epsilon_b\sin^2\chi)-\epsilon_a\epsilon_b=0\,.
\end{equation}
We must choose ${\rm Re}\left[q_c\right]>0$ for localization of energy to the
metal--CTF interface. Let us note parenthetically that $\epsilon_c$ does
not enter our analysis.

The boundary conditions at the interface $z=0$ lead to the two equations
\begin{equation}
\displaystyle{
\left.
\begin{array}{l}
A_c=A_m\\[5pt]
A_c \frac{-iq_c(\epsilon_a\cos^2\chi+\epsilon_b\sin^2\chi)+\vkap(\epsilon_a-\epsilon_b)\sin\chi\cos\chi
}{\vkap^2-(\epsilon_a\cos^2\chi+\epsilon_b\sin^2\chi)} = A_m\frac{i\epsilon_m}{q_m}
 \end{array}\label{eqn:BCs}
 \right\}
 }\,,
\end{equation}
which yield the dispersion equation [11]
\begin{equation}
\displaystyle{
\epsilon_m\vkap^2
+iq_m\vkap(\epsilon_a-\epsilon_b)\sin\chi\cos\chi+
(q_mq_c-\epsilon_m)(\epsilon_a\cos^2\chi+\epsilon_b\sin^2\chi)=0}
\label{eqn:SWP}
\end{equation}
for SWP.

Equation~\r{eqn:SWP} was solved numerically after choosing $\epsilon_m=-56+21i$
(typ., aluminum at $\lambdao=633$~nm [2]) and choosing titanium oxide
as the material of which the CTF is made. At 633-nm wavelength, empirical
relationships have been determined for titanium--oxide CTFs by Hodgkinson
{\em et al.} [8] as
\begin{eqnarray}
\label{eqH1}
\epsilon_a&=&\left[1.0443+2.7394\left(\frac{\chi_v}{\pi/2}\right)-1.3697\left(\frac{\chi_v}{\pi/2}\right)^2 \right]^2\,,\\[5pt]
\epsilon_b&=&\left[1.6765+1.5649\left(\frac{\chi_v}{\pi/2}\right)-0.7825\left(\frac{\chi_v}{\pi/2}\right)^2\right]^2 \,,
\end{eqnarray}
and
\begin{eqnarray}
\tan\chi&=&2.8818\tan\chi_v\,, \label{eqH4}
\end{eqnarray}
where $\chi_v$ and $\chi$ are in radian.  We must caution that the foregoing expressions are applicable to CTFs
produced by one particular experimental apparatus, but may have to be modified for  CTFs produced by other researchers on
different apparatuses.

\smallskip
\begin{center}
\noindent {Table 1.} Dependence of $\vkap$ on $\chi_v$ for
SWP at 633-nm wavelength \\ on a planar interface of aluminum and
a titanium--oxide CTF.
 \\
\bigskip
\begin{tabular}{|c||c|c|}
\hline
$\chi_v$ &    ${\rm Re}[\vkap]$ & ${\rm Im}[\vkap]$ \\
\hline
$5^\circ$ &   $1.2619$ & $0.0179$\\
$20^\circ$ &   $1.8506$  & $0.0174$ \\
$45^\circ$ &    $2.3244$ & $0.0326$ \\
$60^\circ$ &     $2.4688$ & $0.0410$ \\
$90^\circ$  &  $2.5783$ & $0.0488$\\
 \hline
 \end{tabular}
\end{center}
\smallskip

Table 1 shows the computed values of the relative wavenumber $\vkap$ for
SWP on the chosen metal--CTF interface. Clearly, as $\chi_v$ increases towards
$90^\circ$, both the real and the imaginary parts of $\vkap$ increase. This means
that an increase in the vapor incidence angle
\begin{itemize}
\item[(i)] reduces the magnitude of the phase velocity and
\item[(ii)] increases the attenuation
\end{itemize}
of the surface wave. Thus, a high value of $\chi_v$ is inimical to long--range
SWP.\\

\noindent{\bf 3 Modified Kretschmann Configuration}

In conformance with the Kretschmann configuration [9] for launching surface plasmon polaritons,
the half--space $z\leq 0$ is occupied by a homogeneous, isotropic, dielectric
material described by the relative permittivity scalar $\eps_1$. Dissipation in this material
is considered to be negligible and its refractive index $n_1=\sqrt{\eps_1}$
is real--valued and positive. The laminar region
 $0 \leq z\leq L_{m}$ is occupied by a metal with relative permittivity
 scalar $\epsmet$. A CTF of relative permittivity dyadic $\=\eps_{\,ctf}$ occupies the region
 $L_{m} \leq z \leq L_\Sigma=L_{m}+L_{ctf}$. Finally,   the half--space
 $z\geq L_\Sigma$ is taken to be occupied by an isotropic, nondissipative
 dielectric material with relative permittivity $\epsilon_2=n_2^2>0$. This modified
 Kretschmann configuration is in accordance with practical considerations for
 launching surface waves [10].

A $p$--polarized plane
wave, propagating in the half--space
$z \leq 0$ at an angle $\theta_1\in[0,\pi)$ to the $z$ axis   in the $xy$ plane, is incident on the metal--coated CTF.
The electromagnetic field phasors associated
with the incident plane wave are represented as [7]
\begin{equation}
\left.\begin{array}{l}
\Einc= (-\ux\cos\theta_1+\uz\sin\theta_1) \,\exp\left[i(\kappa x+
n_1 z\cos\theta_1)\right]
\\[5pt]
\Hinc= -n_1\etao^{-1}\,\uy \,\exp\left[i(\kappa x+
n_1 z\cos\theta_1)\right]
\end{array}\right\}
\, , \qquad z \leq 0
\, ,
\end{equation}
where
\begin{equation}
\kappa =
\ko n_1\sin\theta_1\,.
\end{equation}
The reflected electromagnetic field phasors are expressed as
\begin{equation}
\left.\begin{array}{l}
\Erefl= \rp(\ux\cos\theta_1+\uz\sin\theta_1) \,\exp\left[i(\kappa x-
n_1 z\cos\theta_1)\right]
\\[5pt]
\Hrefl=-\rp\, n_1\etao^{-1}\,\uy \,\exp\left[i(\kappa x-
n_1 z\cos\theta_1)\right]
\end{array}\right\}
\, , \qquad z \leq 0
\, ,
\end{equation}
and the transmitted electromagnetic field phasors  as
\begin{equation}
\left.\begin{array}{l}
\Etr= \tp (-\ux\cos\theta_2+\uz\sin\theta_2) \,\exp\left\{i\left[\kappa x+
n_2 \left(z-L_\Sigma\right)\cos\theta_2\right]\right\}
\\[5pt]
\Htr= -\tp\, n_2\etao^{-1}\,\uy \, \exp\left\{i\left[\kappa x+
n_2 \left(z-L_\Sigma\right)\cos\theta_2\right]\right\}
\end{array}\right\}
\, , \qquad z \geq L_\Sigma
\, ,
\end{equation}
where $n_2\sin\theta_2=n_1\sin\theta_1=\vkap$.

The reflection coefficient   $\rp$ and the transmission
coefficient  $\tp$ have to be determined by the solution of
a boundary--value problem [7].
It suffices to state here that the  matrix equation
\begin{equation}
\left[\begin{array}{l} \tp\\ 0 \end{array}\right]=
[\=K^{(2)}]^{-1}\cdot\exp\left({i[\=P_{\,ctf}]L_{ctf}}\right)\cdot
\exp\left({i[\=P_{\,m}]L_{m}}\right)\cdot[\=K^{(1)}]\cdot
\left[\begin{array}{l} 1\\ \rp\end{array}\right]\,
\label{finaleq}
\end{equation}
emerges,
wherein the following 2$\times$2 matrixes are employed:
\begin{equation}
[\=K^{(\ell)}] =
\left[ \begin{array}{cc}
-\cos\theta_\ell &\qquad\cos\theta_\ell\\[5pt]
-n_\ell\,\etao^{-1} &\qquad-n_\ell\,\etao^{-1}
\end{array}\right]\,,\quad \ell\in\left\{1,2\right\}\,,
\end{equation}
\begin{equation}
[\=P_{\,m}]=\left[ \begin{array}{cc}
0 &\qquad \omega\muo-\frac{\kappa^2}{\omega\epso\epsmet}
\\[5pt]
\omega\epso\epsmet & \qquad 0
\end{array}\right]\,,
\end{equation}
\begin{equation}
[\=P_{\,ctf}]=\left[ \begin{array}{cc}
\frac{\kappa(\epsilon_a-\epsilon_b)\sin\chi\cos\chi}
{\epsilon_a\cos^2\chi+\epsilon_b\sin^2\chi}
&\qquad \omega\muo-\frac{\kappa^2}{\omega\epso(\epsilon_a\cos^2\chi+\epsilon_b\sin^2\chi)}
\\[5pt]
\omega\epso\frac{\epsilon_a\epsilon_b}
{\epsilon_a\cos^2\chi+\epsilon_b\sin^2\chi}
&\qquad \frac{\kappa(\epsilon_a-\epsilon_b)\sin\chi\cos\chi}
{\epsilon_a\cos^2\chi+\epsilon_b\sin^2\chi}
\end{array}\right]\,.
\end{equation}
The solution of \r{finaleq} yields
the reflection and transmission coefficients.
The principle of conservation of energy mandates
the constraint
\begin{equation}
\vert\rp\vert^2 +\left(\frac{n_2}{n_1}\right)\,
\left(\frac{{\rm Re}[\cos\theta_2]}{\cos\theta_1}\right)\vert\tp\vert^2
\leq1\,,
\end{equation}
the inequality turning to an equality only in the
absence of dissipation in the region $0<z<L_\Sigma$.

In order to study the excitation of a surface wave at the metal--CTF interface,
the absorbance
\begin{equation}
A_p=1-\left\{
\vert\rp\vert^2 +\left(\frac{n_2}{n_1}\right)\,
\left(\frac{{\rm Re}[\cos\theta_2]}{\cos\theta_1}\right)\vert\tp\vert^2
\right\}
\end{equation}
has to be plotted against the angle $\theta_1$. This was done
for  $L_m=10$~nm and $L_{ctf}=1000$~nm.
As in Section 2, $\epsilon_m=-56+21i$ was chosen,
along with $\eps_{a,b}$ and $\chi$ as specified by \r{eqH1}--\r{eqH4}. Whereas
 $\epsilon_1=9$ was chosen to ensure that existence of a critical
 angle for total reflection in the absence of the 10-nm--thick metal film,
 $\epsilon_2=5$ was set for a lesser degree of constitutive contrast with the CTF.

Figure~2 shows plots of $A_p$ vs. $\theta_1$ for the same
five values of $\chi_v$ as in Table~1. For comparison, plots
of $\vert\rp\vert^2$ vs. $\theta_1$ in the absence of
the metal film are also shown in this
figure, in order to establish the critical (minimum) values of $\theta_1$
for total reflection. For $\theta_1$ greater than the critical
angle for a specific $\chi_v$, we see very high $A_p$--peaks in Figure~2.

\smallskip
\begin{center}
\noindent {Table 2.} Dependence of $\theta_1$ and $\vkap$ of the
rightmost peaks in Figure~2 on $\chi_v$. \\
\bigskip
\begin{tabular}{|c||c|c|}
\hline
$\chi_v$ &    $\theta_1$ & $\vkap=n_1\sin\theta_1$ \\
\hline
$5^\circ$ &   $25.782^\circ$ & $1.3037$\\
$20^\circ$ &   $39.515^\circ$  & $1.9088$ \\
$45^\circ$ &    $53.247^\circ$ & $2.4037$ \\
$60^\circ$ &     $58.185^\circ$ & $2.5493$ \\
$90^\circ$  &  $62.168^\circ$ & $2.6530$\\
 \hline
 \end{tabular}
\end{center}
%\smallskip

%%%%%%%%%%  Figure 2 begins %%%%%%%%%%%%
%\smallskip
\begin{center}
\begin{figure}[!htb]
\centering \psfull
\epsfig{file=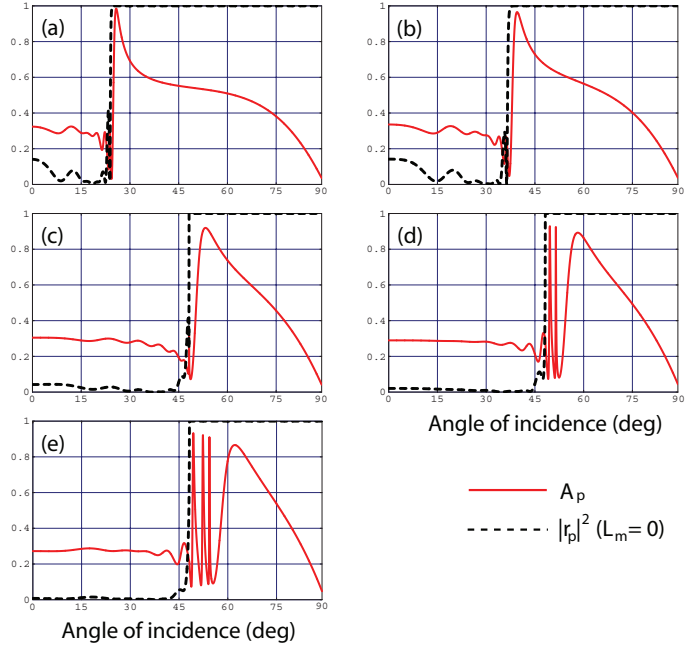, width=10cm}
%\bigskip
\caption{
{\small Solid lines represent the absorbance $A_p$ against the incidence
incidence angle $\theta_1$, when $\epsilon_1=9$, $\epsilon_2=5$,
$\epsilon_m=-56+21i$, $L_m=10$~nm, $L_{ctf}=1000$~nm,
and $\eps_{a,b}$ and $\chi$ are specified by \r{eqH1}--\r{eqH4}.
Dashed lines represent $\vert\rp\vert^2$
against $\theta_1$ when $L_m=0$.
(a) $\chi_v=5^\circ$,
(b) $\chi_v=20^\circ$,
(c) $\chi_v=45^\circ$,
(d) $\chi_v=60^\circ$,
(e) $\chi_v=90^\circ$.
\label{Fig:results}
}}
\end{figure}
\end{center}
\smallskip
%%%%%%%%%%  Figure 2 ends  %%%%%%%%%%%%

Values of $\theta_1$ for  the rightmost peaks of $A_p$ against
$\chi_v$ in Figure~2 are
provided in Table~2, along with the corresponding values of
$\vkap=n_1\sin\theta_1$. The values of $\vkap$ in Table~2
are quite close to the real parts of those in Table~1, thereby indicating
that the rightmost $A_p$--peaks in Figure 1 can be attributed to
the excitation of surface waves at the planar metal--CTF interface [2]. Not
surprisingly, the value of $\theta_1$ needed to excite a surface
wave in the modified Kretschmann configuration increases
as the vapor incidence angle $\chi_v$ gets closer to $90^\circ$.\\

\noindent{\bf 4 Concluding Remarks}

Right from the time of Faraday [12], morphology is widely known to play a key role in the optical
responses of thin films [13-15]. More recently, Hodgkinson and Wu made a comprehensive proposal to
regard columnar thin films as equivalent to biaxial crystals [6], which functional analogy can be
extended to sculptured thin films [7]. Therefore, it is appropriate that the excitation of surface waves
at the planar interfaces of isotropic dielectric materials [10] and metals with CTFs be influenced by
morphology. In this communication, we have shown that the selection of a higher vapor deposition angle
when growing the CTF, with consequent influence on the morphology of the
CTF,
will lead to a surface wave at a planar metal--CTF interface with phase velocity of lower magnitude and
shorter propagation range; concomitantly, a higher incidence angle will be required to excite that
surface wave in a modified Kretschmann configuration.

{\small \noindent{\bf Acknowledgment.} This communication is dedicated to Prof. Prasad Khastgir, a great teacher of physics.}

\smallskip
\noindent{\bf References}
{\small
\begin{enumerate}

\item%1
Zenneck J, \emph{Ann. Phys. Lpz.\/} 23 (1907) 846.

\item%2
Mansuripur M, Li L,
\emph{OSA Opt. Photon. News\/} 8 (1997) 50 (May issue).

\item%3
Pitarke J M,  Silkin V M,  Chulkov E V,  Echenique P M,
\emph{Rep. Prog.  Phys.\/} 70  (2007) 1.

\item%4
Singh J, Thyagarajan K,
\emph{Opt. Commun.\/} 85 (1991) 397.

\item%5
Mihalache D, Baboiu D-M, Ciumac M, Torner L, Torres J P,
\emph{Opt. Quant. Electron.\/} 26 (1994) 857.

\item%6
Hodgkinson I J, Wu Q h,
\emph{Birefringent thin films and polarizing elements.\/}
(World Scientific, Singapore), 1997.

\item%7
Lakhtakia A,  Messier R, \emph{Sculptured thin films:
Nanoengineered morphology and optics.\/} (SPIE Press, Bellingham, WA, USA), 2005,
Chap. 7.

\item%8
Hodgkinson I J, Wu Q h, Hazel J,
\emph{Appl. Opt.\/} 37 (1998) 2653.

\item%9
Kretschmann E,  Raether H,
\emph{Z. Naturforsch. A\/} 23 (1968) 2135.

\item%10
Simon H J, Mitchell D E,   Watson J G,
\emph{Am. J. Phys.\/} 43 (1975) 630.

\item%11
Polo Jr J A, Nelatury S R, Lakhtakia A,
\emph{J. Nanophoton.\/} 1 (2007) 013501.

\item%12
Faraday M,
\emph{Phil. Trans. R. Soc. Lond.\/} 147 (1857) 145.

\item%13
Heavens O S,
\emph{Optical properties of thin solid films.\/} Dover Publications, Mineola, NY, USA),
1965.

\item%14
Ward L,
\emph{The optical constants of bulk materials and films.\/} (Institute of
Physics Publishing, Bristol, United Kingdom), 1994.

\item%15
Rancourt J D,
\emph{Optical thin films: User handbook.\/}
(SPIE Press, Bellingham, WA, USA), 1996.

\end{enumerate}

\end{document}